\documentclass[twocolumn]{aastex63}

\usepackage{amsmath}

\received{March 13, 2020}
\revised{\today}

\submitjournal{ApJ}

\shorttitle{Wind-Driven Model for Peculiar Transients AT2018cow and iPTF14hls}
\shortauthors{UNO \& MAEDA}

\graphicspath{{./}{figures/}}

\begin{document}

\title{A Wind-Driven Model: Application to Peculiar Transients AT2018cow and iPTF14hls}

\correspondingauthor{KOHKI UNO}
\email{k.uno@kusastro.kyoto-u.ac.jp}

\author{KOHKI UNO}
\affiliation{Department of Astronomy, Kyoto University, Kitashirakawa-Oiwake-cho, Sakyo-ku, Kyoto, 606-8502, Japan}

\author{KEIICHI MAEDA}
\affiliation{Department of Astronomy, Kyoto University, Kitashirakawa-Oiwake-cho, Sakyo-ku, Kyoto, 606-8502, Japan}

\begin{abstract}
We propose a wind-driven model for peculiar transients, and apply the model to AT2018cow and iPTF14hls. In the wind-driven model, we assume that a continuous outflow like a stellar wind is injected from a central system. While these transients have different observational properties, this model can explain their photometric properties which are not reproduced by a supernova-like instantaneous explosion. Furthermore, the model predicts characteristic spectral features and evolution, which are well in line with those of AT2018cow and iPTF14hls. Despite the different observational properties, the wind model shows that they have some common features; the large mass-loss rates (up to $\sim 20M_{\odot}{\rm ~yr^{-1}}$ for AT2018cow and $\sim 75M_{\odot}{\rm ~yr^{-1}}$ for iPTF14hls), and the characteristic radii of $\sim 10^{13}{\rm ~cm}$ for the launch of the wind. It would indicate that both may be related to events involving a red super giant (RSG), in which the RSG envelope is rapidly ejected by an event at a stellar core scale. On the other hand, the main differences are the kinetic energies, the total ejected mass, and time scales. We then suggest that iPTF14hls may represent a dynamical common-envelope evolution induced by a massive binary system ($\sim 120M_{\odot}+100M_{\odot}$). AT2018cow may be either a tidal disruption event of a low-mass RSG by a black hole (BH), or a BH-forming failed supernova.
\end{abstract}


\section{Introduction} \label{sec:1}

In recent years, new classes of astronomical transients have been discovered, thanks to improvement in observational instruments and operation of new generation surveys such as Pan-STARRS \citep{2002SPIE.4836..154K}, PTF \citep{2009PASP..121.1395L}, ASSA-SN \citep{2014AAS...22323603S}, and ZTF \citep{2018ATel11266....1K}. Some transients have peculiar light curves and/or spectral evolution, comprehensive understanding of which has not been reached by any existing models. AT2018cow \citep{2018ApJ...865L...3P} and iPTF14hls \citep{2017Natur.551..210A} are among these enigmatic transients, whose origins have not been identified yet.

AT2018cow is a fast and luminous blue transient, discovered by ATLAS on MJD 58285.44141 \citep{2018ApJ...865L...3P}. It showed high luminosity (up to $L \sim 10^{44} {\rm ~erg~s^{-1}}$) which exceeds those of superluminous supernovae(SNe), a rapidly declining luminosity roughly following a power law, and a recessing photospheric radius from the beginning. Furthermore, it showed a very high velocity $(v \sim 0.1 c)$, where $c$ is the speed of light, at the beginning \citep{2019MNRAS.484.1031P}. These features are different from those seen in SNe. It also showed characteristic spectral evolution. In the optical/UV wavelengths, broad emission lines of hydrogen and helium developed after $\sim 15$ days from the discovery, and they showed redshifts of $\sim 3000{\rm ~km~s^{-1}}$. After that, the lines evolved blueward, and eventually developed sharp peaks around the rest wavelengths \citep{2019MNRAS.484.1031P, 2019MNRAS.487.2505K}. It also showed strong radio and X-ray emissions \citep{2019ApJ...871...73H, 2018MNRAS.480L.146R, 2019ApJ...872...18M, 2020MNRAS.491.4735B}. Based on these observational properties, some models have been proposed, including an electron-capture collapse \citep{2019MNRAS.487.5618L}, a Tidal Disruption Event \citep[TDE, ][]{2019MNRAS.484.1031P, 2019MNRAS.487.2505K}, a common envelope jet \citep{2019MNRAS.484.4972S}, a magnetar formation \citep{2020ApJ...888L..24M}, and a fallback accretion following a collapse of a blue supergiant \citep{2019ApJ...872...18M}. However, most, if not all, of the proposed models aim at explaining its energetics, luminosity, or time scale. The origin of the most peculiar observational features in the time evolution, as described above, remains unanswered.

iPTF14hls was discovered by the iPTF survey on MJD 56922.53 \citep{2017Natur.551..210A}. It was classified as a typical Type IIP SN \citep{1997ARA&A..35..309F} at the beginning. However, it turned out to keep high brightness for almost 2 years \citep{2017Natur.551..210A}. Although snapshot spectra of iPTF14hls were very similar to Type IIP SNe (e.g., showing hydrogen lines with the P-Cygni profile), its evolution was too slow; it showed line velocities ($\sim 4000{\rm ~km~s^{-1}}$) and the color nearly constant over time. After 2 years, the luminosity started to show a decrease, and the spectra finally turned into nebular ones \citep{2019A&A...621A..30S}. While its long timescale itself is peculiar, what is indeed the most difficult to understand is this combination of the (nearly) constant color (temperature) and the constant line velocities. Some models \citep[e.g., ][]{2018ApJ...863..105W, 2018MNRAS.477...74A,2018A&A...610L..10D, 2019MNRAS.485L..83Q, 2019JHEAp..22....5L, 2019MNRAS.482.4233G, 2019MNRAS.488.5854G} have been proposed, but mostly dealing with the light curve behavior.

Explosions like SNe produce homologously expanding ejecta, with the monotonically increasing physical scales and decreasing density and optical depth. This combination never explains the peculiar time evolution seen in AT2018cow and iPTF14hls as described above. The homologous expansion predicts that the photospheric radius increases initially (unlike AT2018cow). If the luminosity stays nearly constant (within a factor of a few), it must show either decreasing temperature or decreasing line velocities (unlike iPTF14hls).

These peculiar properties suggest that these systems might be described as a (stellar) wind (i.e., a continuous input of the mass and the energy from the inner engine) rather than an SN-like explosion (i.e. an instantaneous explosion). Indeed, \citet{2020MNRAS.491.1384M} suggested such a model for iPTF14hls based on a phenomenological argument (Section \ref{subsec:3.2} for more details). In this paper, we present a physically-motivated model for the `wind-driven' explosion. We apply the model to AT2018cow and iPTF14hls, and show that their light curves and the evolution of the photosphere (i.e. color) can be explained within the same context. Furthermore, we investigate the details of the spectral line formation process, and find that the model predictions are perfectly in line with the characteristic line properties and the spectral evolution for both transients.

The paper is structured as follows. In Section \ref{sec:2}, we introduce an analytical setup of the wind-driven model, under the assumption of the steady state. In Section \ref{sec:3}, we apply the model to AT2018cow and iPTF14hls, and estimate the mass-loss rates and other wind properties using their photometric data. We further discuss the properties of spectral line formation and its evolution, and the model predictions here are compared with the spectroscopic properties of AT2018cow and iPTF14hls. Based on the derived properties of the wind, we discuss possible origins of these transients in Section \ref{sec:4}. The paper is closed in Section \ref{sec:5} with conclusions.

When we were finalizing this manuscript, \citet{2020ApJ...894....2P} presented their new work in which they independently treated the same configuration and already derived most of the contents described in Section \ref{sec:2}. They also discussed applicability of their model to AT2018cow. In this paper, we investigate further details, and apply the wind-driven model to both of AT2018cow and iPTF14hls, which allows the discussion about the similarities between these transients with different observational features, as well as key differences. This is important to further discuss their possible origins as we do in the present work. Furthermore, we discuss the detail of the spectral line formation, which has been missing in the previous works.

\section{WIND-DRIVEN MODEL} \label{sec:2}

The basic formalism described here has been independently derived by the earlier work by \citet{2020ApJ...894....2P}. We note that additional processes (e.g., recombination and spectral formation) are newly discussed in the present work.

\begin{figure}
\plotone{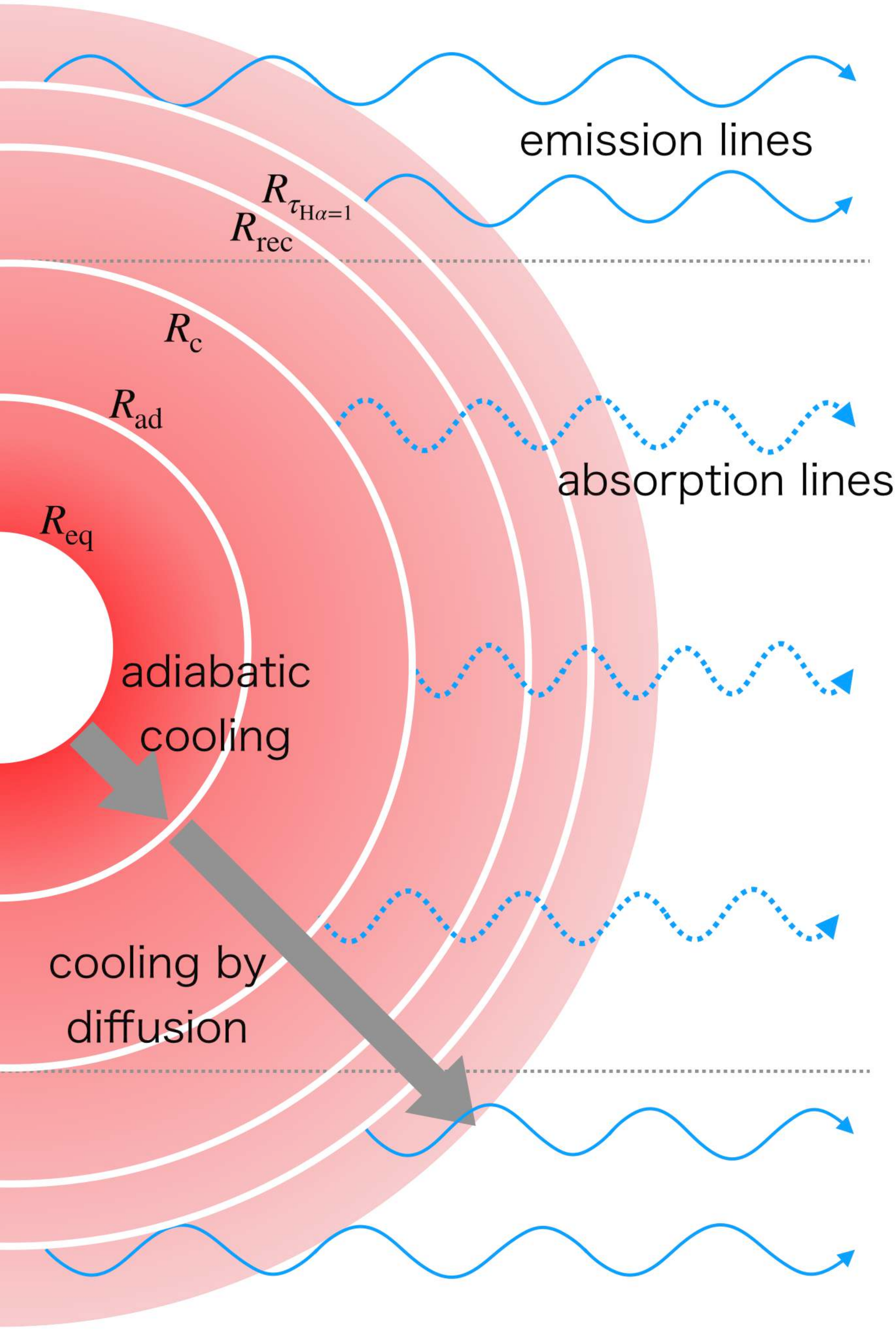}
\caption{Schematic picture of the wind-driven model. Considering the Sobolev optical depth (see Appendix \ref{sec:appendixA}), we define a radius where $\rm H\alpha$ is formed $(R_{\rm \tau_{\rm H\alpha}=1})$. In this example, the color radius $(R_{\rm c})$ becomes the photospheric radius. The figure shows a concept of the line formation, including where the absorption (dotted lines) and emission (solid) are created.}
\label{fig:fig1}
\end{figure}

In the wind-driven model (see Figure \ref{fig:fig1}), we consider continuous outflows, which are analogous to stellar winds, characterized by the mass loss rate $(\dot{M})$ and the wind velocity $(v)$. Under the assumption of steady states, the density structure of the system is given as follows;
\begin{equation}
    \rho(r) = \frac{\dot{M}}{4\pi r^2 v}.
\end{equation}
Throughout the paper, we assume that the wind velocity is constant as a function of radius for each snapshot, while the velocity at the wind launch can change as a function of time. The effect of the wind acceleration \citep[e.g., ][]{2020MNRAS.491.1384M} is beyond a scope of the present work. We, however, note that it would not affect the main conclusions of the present work. First of all, the model basically uses the photospheric properties in its construction, and therefore the input velocity here can simply be regarded as the value at the photosphere. The possible wind acceleration, therefore, would not affect the derived mass-loss rate. Indeed, the possible wind acceleration would affect only the spectral formation, since this happens above the photosphere. However, even if the wind velocity would be doubled along its trajectory, the characteristic values (i.e., density and optical depths) are changed only by a factor of two. This effect is too small to change the overall spectral properties.

The innermost (equipartition) radius is described as $R_{\rm eq}$, which can be regarded as the position where the wind is launched. In the inner region above $R_{\rm eq}$, matter and photons are coupled up to the radius $R_{\rm ad}$, where $\tau_{\rm s} \approx c/v$ (where $\tau_{\rm s}$ is the optical depth considering electron scattering). The temperature there is decreasing adiabatically. Above this region, the luminosity is roughly constant and the temperature there is determined by photon diffusion. Within the outer region, some characteristic radii, $R_{\rm c}$, $R_{\rm rec}$, $R_{\rm s}$ and $R_{\rm \tau_{\rm H\alpha}=1}$, are defined. $R_{\rm c}$ is the color radius, where $\tau_{\rm eff} \approx 1$ (where $\tau_{\rm eff}$ is the effective optical depth). $R_{\rm rec}$ is the recombination radius. $R_{\rm s}$ is the scattering radius where $\tau_{\rm s}\approx 1$. $R_{\rm \tau_{\rm H\alpha}=1}$ is the ${\rm H\alpha}$ line-forming radius (see Figure \ref{fig:fig1}).

The optical depth for electron scattering is defined as follows;
\begin{equation}
  \tau_{\rm s} = \int_r^{R_{\rm out}}\kappa_{\rm s}\rho(r)dr = \frac{\kappa_{\rm s} \dot{M}}{4\pi v}\left(\frac{1}{r}-\frac{1}{R_{\rm out}}\right),
\end{equation}
where $\kappa_{\rm s}$ is the opacity considering electron scattering ($\kappa_{\rm s} = 0.34{\rm ~cm^2~g^{-1}}$ for the solar composition). $R_{\rm out}$ is defined as the outermost radius, above which $\kappa_{\rm s} = 0$. If the relation $R_{\rm out} \gg r$ holds, $\tau_{\rm s}$ is described as follows;
\begin{equation}
  \tau_{\rm s} = \frac{\kappa_{\rm s} \dot{M}}{4\pi v}\frac{1}{r}.
\end{equation}
The effective optical depth $(\tau_{\rm eff})$, considering not only electron scattering but also absorption processes, is defined as follows;
\begin{equation}
  \tau_{\rm eff} = \int_{r}^{R_{\rm out}} \kappa_{\rm eff} \rho(r) dr,
\end{equation}
where $\kappa_{\rm eff}$ is the effective opacity, given as follows;
\begin{equation}
  \kappa_{\rm eff} = \sqrt{3(\kappa_{\rm s}+ \kappa_{\rm a})\kappa_{\rm a}} \approx \sqrt{3\kappa_{\rm s} \kappa_{\rm a}}\quad(\kappa_{\rm s} \gg \kappa_{\rm a}).
\end{equation}
For the Kramar's opacity, we use $\kappa_{\rm a} = \kappa_{0}\left(\frac{\rho}{\rm{~g~cm^{-3}}}\right) \left(\frac{T}{\rm K}\right)^{-7/2}$ with $\kappa_0 = 2 \times 10^{24} {\rm ~cm^2~g^{-1}}$ \citep{2020ApJ...894....2P}.

We define the innermost radius $(R_{\rm eq})$ as the radius below which equipartition is realized between the internal energy (dominated by radiation) and the kinetic energy; $aT^{4} = \rho v^{2}/2$ where $a$ is the radiation constant. The temperature there, $T_{\rm eq} = T(R_{\rm eq})$, is then described as follows;
\begin{equation}
  T_{\rm eq} = \left(\frac{1}{8\pi a}\right)^{\frac{1}{4}}\dot{M}^{\frac{1}{4}}v^{\frac{1}{4}}R_{\rm eq}^{-\frac{1}{2}}.
\end{equation}

Above $R_{\rm eq}$, the temperature first decreases adiabatically as a function of radius, following the advection by the wind. The outermost radius of this region is defined as $R_{\rm ad}$ \citep[see also ][]{2009MNRAS.400.2070S}. It is defined by $\tau_{\rm s}\approx c/v$, and thus
\begin{equation}
  R_{\rm ad} = \frac{\kappa_{\rm s}}{4\pi c}\dot{M}.
\end{equation}
The temperature structure at $R_{\rm eq} < r < R_{\rm ad}$ is given as follows;
\begin{equation}
\label{adiabatic}
  T(r) = T_{\rm eq}\left(\frac{r}{R_{\rm eq}}\right)^{-\frac{2}{3}}.
\end{equation}
Above $R_{\rm ad}$, the temperature is determined by photon diffusion. The temperature structure is then described as follows;
\begin{equation}
\label{diffusion}
  T(r) = T_{\rm ad}\left(\frac{r}{R_{\rm ad}}\right)^{-\frac{3}{4}},
\end{equation}
where $T_{\rm ad} = T(R_{\rm ad})$ is given as follows;
\begin{equation}
\label{eq:Tad}
\begin{split}
T_{\rm ad} &=T_{\rm eq}\left(\frac{R_{\rm ad}}{R_{\rm eq}}\right)^{-\frac{2}{3}} \\ 
&=\left(\frac{2^{\frac{7}{12}}\pi^{\frac{5}{12}}c^{\frac{2}{3}}}{a^{\frac{1}{4}}\kappa_{\rm s}^{\frac{2}{3}}}\right) \dot{M}^{-\frac{5}{12}} v^{\frac{1}{4}} R_{\rm eq}^{\frac{1}{6}}.
\end{split}
\end{equation}

Using the distribution of density and temperature, we estimate the color (thermalization) radius $R_{\rm c}$;
\begin{equation}
\label{eq:Tc}
    R_{\rm c} = \left(\frac{2^{\frac{125}{33}}\cdot3^{\frac{8}{11}}c^{\frac{7}{33}}a^{\frac{7}{11}}\kappa_{0}^{\frac{8}{11}}\kappa_{\rm s}^{\frac{17}{33}}}{11^{\frac{16}{11}}\pi^{\frac{4}{3}}}\right)\dot{M}^{\frac{4}{3}}v^{-\frac{31}{11}}R_{\rm eq}^{-\frac{14}{33}},
\end{equation}
in case $R_{\rm c} \ll R_{\rm rec}$ (where the recombination radius, $R_{\rm rec}$, is described below).
The formation of the photosphere depends on a relative configuration between $R_{\rm c}$ and $R_{\rm ad}$. 
When $R_{\rm c} < R_{\rm ad}$, the photons emitted at $R_{\rm c}$ are still trapped up to $R_{\rm ad}$. Therefore, the photospheric temperature is determined by $R_{\rm ad}$, and $R_{\rm ad}$ becomes the photospheric radius; $R_{\rm ph}$ is then determined by $\tau_{\rm s}(R_{\rm ph}) = c/v$. On the other hand, if $R_{\rm c} > R_{\rm ad}$ holds, $R_{\rm c}$ becomes $R_{\rm ph}$ $(\tau_{\rm eff}(R_{\rm ph}) = 1)$.

An additional physical scale is introduced by the ionization structure. We consider the recombination radius $R_{\rm rec}$, as defined by $T(R_{\rm rec})=T_{\rm rec}$, where $T_{\rm rec}$ is the recombination temperature. In the present work, $T_{\rm rec}$ is taken as $6000 {\rm ~K}$ and $12000{\rm ~K}$ for $\rm H$ and $\rm He$, respectively, as are typical for the density considered in the present work \citep{1996snih.book.....A}. In general $T_{\rm ad} > T_{\rm rec}$ holds, and thus $R_{\rm rec}$ is determined as follows thorough equation \eqref{diffusion};
\begin{equation}
  T_{\rm rec} = T_{\rm ad}\left(\frac{R_{\rm rec}}{R_{\rm ad}}\right)^{-\frac{3}{4}}.
\end{equation}

The luminosity is given by the photon diffusion. Above $R_{\rm ad}$, the flux $F(r)$ must be nearly constant;
\begin{equation}
  L(r) = -\frac{4\pi r^2 a c}{3\kappa_{\rm s} \rho}\frac{\partial}{\partial r}T^{4} \approx {\rm constant}.
\end{equation}
Therefore,
\begin{equation}
\label{eq:L}
\begin{split}
    L(r) = 8.&92\times 10^{43}~{\rm erg~s^{-1}}\\
    &\times\left(\frac{\dot{M}}{M_{\odot}{\rm~ yr^{-1}}}\right)^{\frac{1}{3}}\left(\frac{v}{0.1c}\right)^{2}\left(\frac{R_{\rm eq}}{1\times 10^{13}{\rm ~cm}}\right)^{\frac{2}{3}}.
\end{split}
\end{equation}
For most of the cases, the photosphere is formed above $R_{\rm ad}$. So that this formula can be used.

If the color temperature, described as $T_{\rm c} = T(R_{\rm c})$, and $T_{\rm ad}$ are close to the recombination temperature, we are not able to use the approximation $R_{\rm rec} \gg R_{\rm c}$ and $R_{\rm rec} \gg R_{\rm ad}$. Then, we need to take the effects of the recombination radius into account, i.e., it is necessary to recalculate $\tau_{\rm s}$ and $\tau_{\rm eff}$ by adding $-1/R_{\rm rec}$, which we have ignored so far. Then we need to solve the following three relations.
First, the relation $\tau_{\rm s} (R_{\rm ad}) = c/v$ must be satisfied. With the temperature structures (the relations \eqref{adiabatic} and \eqref{diffusion}), it is described as follows;
\begin{equation}
\label{eq:close1}
\begin{split}
  R_{\rm ad} = (8\pi a)^{3}T_{\rm rec}^{12}\left(\frac{1}{R_{\rm ad}} - \frac{4\pi c}{\kappa_{\rm s}\dot{M}}\right)^{-9}\\
 \times\dot{M}^{-3}v^{-3}R_{\rm eq}^{-2}.
\end{split}
\end{equation}
Second, the condition $\tau_{\rm eff} (R_{\rm c}) = 1$ is described as follows;
\begin{equation}
\label{eq:close2}
\begin{split}
  1 = \frac{8}{11}\pi^{-\frac{5}{6}}\sqrt{3\kappa_{\rm s}\kappa_{0}}a^{\frac{2}{3}}
  \left( T_{\rm c}^{\frac{11}{12}}-T_{\rm rec}^{\frac{11}{12}} \right)\\
  \times\dot{M}^{\frac{5}{6}} v^{-\frac{13}{6}}R_{\rm eq}^{-\frac{4}{9}}R_{\rm ad}^{-\frac{2}{9}}.
\end{split}
\end{equation}
Third, the luminosity is given as follows;
\begin{equation}
\label{eq:close3}
\begin{split}
  L = \frac{2\pi c}{\kappa_{\rm s}}v^2R_{\rm eq}^{\frac{2}{3}}R_{\rm ad}^{\frac{1}{3}}.
\end{split}
\end{equation}

In the wind-driven model, we can compute two observables (luminosity and photospheric temperature) from three input parameters ($R_{\rm eq}$, $\dot{M}$, and $v$). Conversely, from observational data of luminosity and photospheric temperature, we can estimate (or give constraints on) these parameters. However, using only two observables would not give a unique solution. Practically, we can use another observational information (e.g., line velocity) to close the relations, the examples of which are given in Section \ref{sec:3}.

\section{APPLICATIONS TO THE OBSERVED TRANSIENTS} \label{sec:3}

\subsection{AT2018cow}

Using the relations we derived in Section \ref{sec:2}, we can calculate $R_{\rm eq}(t)$, $\dot{M}(t)$ and $v(t)$ under the wind-driven model from the observational data \citep{2019MNRAS.484.1031P}, i.e., luminosity $L(t)$ and observed photospheric temperature $T_{\rm ph}(t)$. From the observationally inferred photospheric radius (from $L$ and $T_{\rm ph}$), the initial velocity of AT2018cow must be $\sim 0.1c$. This constraint can be used to derive a unique solution for $R_{\rm eq}$, $\dot{M}$, and $v$, at the initiation of the outflows. However, after that, the evolution of $v$ is not clear. As a rational approximation, we assume that $R_{\rm eq}$ is constant over time, which is then fixed by the above information.

At the initiation of the event, $L=3.4\times10^{44}{\rm ~erg~s^{-1}}$, $T_{\rm ph} = 31390{\rm ~K}$, and $v = 0.1c$. From the equations \eqref{eq:Tad} and \eqref{eq:L}, $R_{\rm eq}$ is then derived as $1.7\times 10^{13}{\rm ~cm}$, and this radius is fixed for subsequent evolution. In addition, the relation $R_{\rm c} < R_{\rm ad}$ holds in the early phase, and thus $R_{\rm ph} = R_{\rm ad}$ and $T_{\rm ph} = T(R_{\rm ad})$. Then, using the relations, \eqref{eq:Tad} and \eqref{eq:L}, $\dot{M}(t)$ and $v(t)$ are derived as follows;
\begin{equation}
  \begin{split}
    v(t) = 3.&00\times10^9{\rm~cm~s^{-1}}\\
    &\times\left(\frac{T_{\rm ph}(t)}{31390{\rm K}}\right)^{\frac{4}{11}}\\
    &\times\left(\frac{L(t)}{3.4\times10^{44}{\rm ~erg~s^{-1}}}\right)^{\frac{5}{11}}, ~\rm and
\end{split}
\end{equation}
\begin{equation}
  \begin{split}
    \dot{M}(t) = 1&9.4 M_{\odot}{\rm ~yr^{-1}}\\
    &\times\left(\frac{v(t)}{3\times10^9{\rm ~cm~s^{-1}}}\right)^{-\frac{1}{2}}\left(\frac{T_{\rm ph}(t)}{31390{\rm ~K}}\right)^{-2}\\
    &\times\left(\frac{L(t)}{3.4\times10^{44}{\rm ~erg~s^{-1}}}\right)^{\frac{1}{2}}.
  \end{split}
\end{equation}

\begin{figure}
\epsscale{1.17}
\plotone{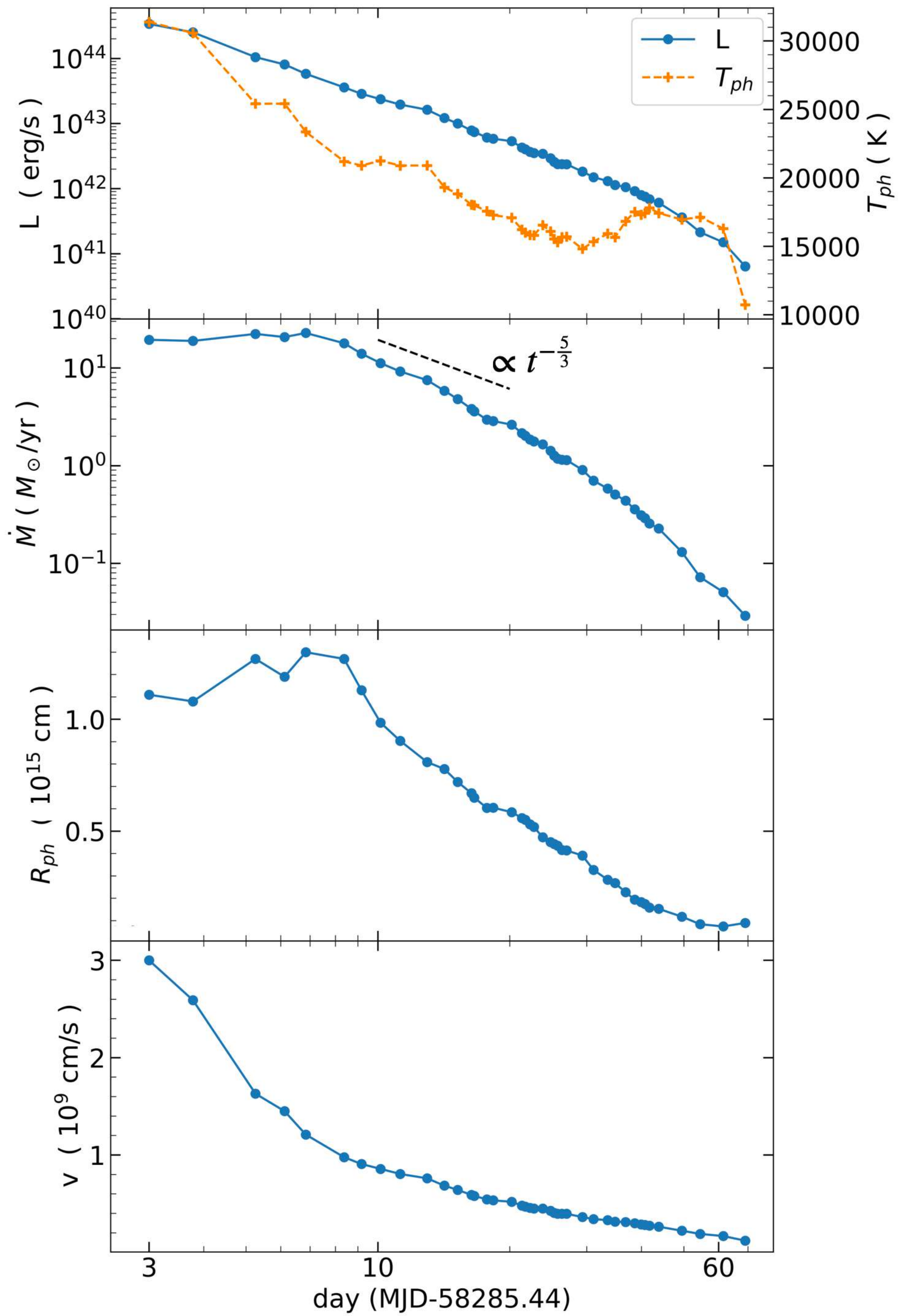}
\caption{The top panel shows the bolometric light curve of AT2018cow (blue circles; left axis) from \cite{2019MNRAS.484.1031P} and the evolution of the photospheric temperature (orange crosses; right axis). The second panel shows the evolution of the derived mass-loss rate. The black dashed line shows a power law with the index of $-5/3$. The third panel shows the estimated photospheric radius evolution. The bottom panel shows the estimated velocity evolution.}
\label{fig:fig2}
\end{figure}

After a few days, the relation between the characteristic radii turns out to change to $R_{\rm c} > R_{\rm ad}$, and thus we use $R_{\rm ph} = R_{\rm c}$ and $T_{\rm ph} = T(R_{\rm c})$. Then, using expressions \eqref{eq:Tc} and \eqref{eq:L}, $\dot{M}(t)$ and $v(t)$ are derived as follows;
\begin{equation}
  \begin{split}
    v(t) = 9.&86\times10^8{\rm ~cm~s^{-1}}\\
    &\times\left(\frac{T_{\rm ph}(t)}{21200{\rm ~K}}\right)^{\frac{11}{70}}\\
    &\times\left(\frac{L(t)}{3.6\times10^{43}{\rm ~erg~s^{-1}}}\right)^{\frac{11}{35}}, ~\rm and
\end{split}
\end{equation}
\begin{equation}
  \begin{split}
    \dot{M}(t) = &18.0 M_{\odot}{\rm yr^{-1}}\\
    &\times\left(\frac{v(t)}{9.86\times10^8{\rm ~cm~s^{-1}}}\right)\left(\frac{T_{\rm ph}(t)}{21200{\rm ~K}}\right)^{-\frac{11}{10}}\\
    &\times\left(\frac{L(t)}{3.6\times10^{43}{\rm ~erg~s^{-1}}}\right)^{\frac{4}{5}}.
  \end{split}
\end{equation}
Figure \ref{fig:fig2} shows the evolution of $\dot{M}$, $v$, and $R_{\rm ph}$ as we have derived.
In the wind-driven model, $\dot{M}$ is roughly constant for $\sim 10$ days after the initiation. After that, $\dot{M}$ decreases following a power law as a function of time (see Figure \ref{fig:fig2}). Interestingly, the power law behavior with the index of $-5/3$ is found, which is the typical mass accretion rate evolution for fallback of materials onto a central compact object (e.g., TDE or failed SN). Therefore, it points to a possibility that a power source of AT2018cow may be accretion onto a compact object (see Section \ref{sec:4}). 

By a rough application of a similar (basically the same) model to AT2018cow, \citet{2020ApJ...894....2P} reached to the similar conclusion, as we confirm here. Note that the behavior in the first $\sim 10$ days is different. This is due to a difference in the detail of the model. While \citet{2020ApJ...894....2P} assumed that $v$ is constant over time, we allow the evolution of $v$ under the constraint given by the initial condition.

Within the wind-driven model, very strong outflows (over $20M_{\odot}{\rm ~yr^{-1}}$) immediately after the initiation of the explosive event are required. After $\gtrsim 10$ days, the estimated mass-loss rate decreases to a few $M_{\odot}{\rm ~yr^{-1}}$, and the wind velocity becomes as low as $4000{\rm ~km~s^{-1}}$. Integrating the estimated mass-loss rate and the kinetic power over time, we estimate that the total ejected mass $(M_{\rm total})$ is $0.68M_{\odot}$ and total kinetic energy $(E_{\rm kin,total})$ is $1.7\times 10^{51} {\rm ~erg}$. Note that the cumulative kinetic energy exceeds $10^{51}{\rm ~erg}$ already at $\sim 4$ days (see Figure \ref{sec:3}). Thus, the outflows immediately after the initiation contain most of the total kinetic energy.

\begin{figure}
\epsscale{1.17}
\plotone{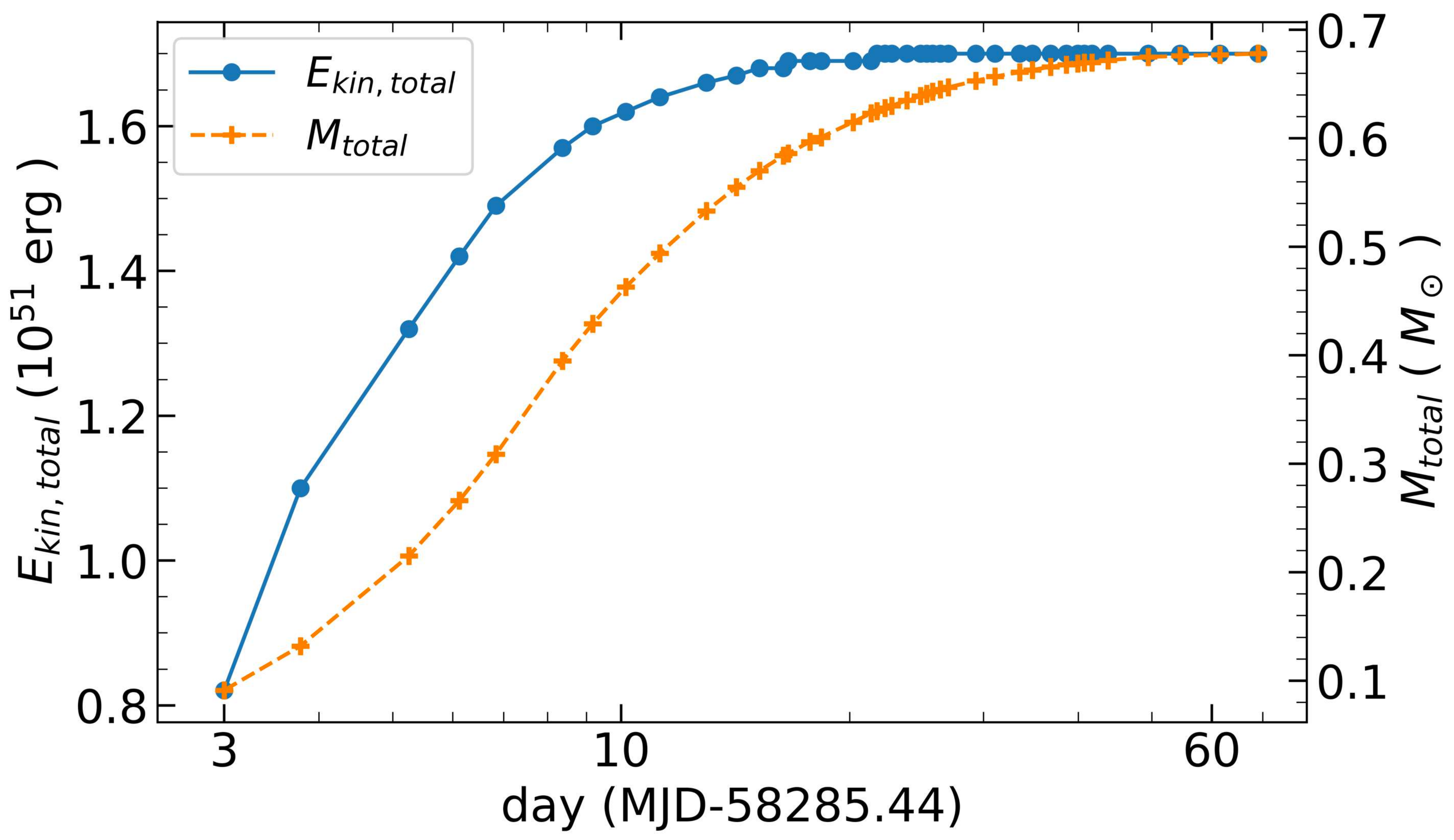}
\caption{The evolution of the total kinetic energy (blue circles; left axis) and the total ejected mass (orange crosses; right axis) in the wind-driven model for AT2018cow.}
\label{fig:fig3}
\end{figure}

\begin{figure}
\epsscale{1.17}
\plotone{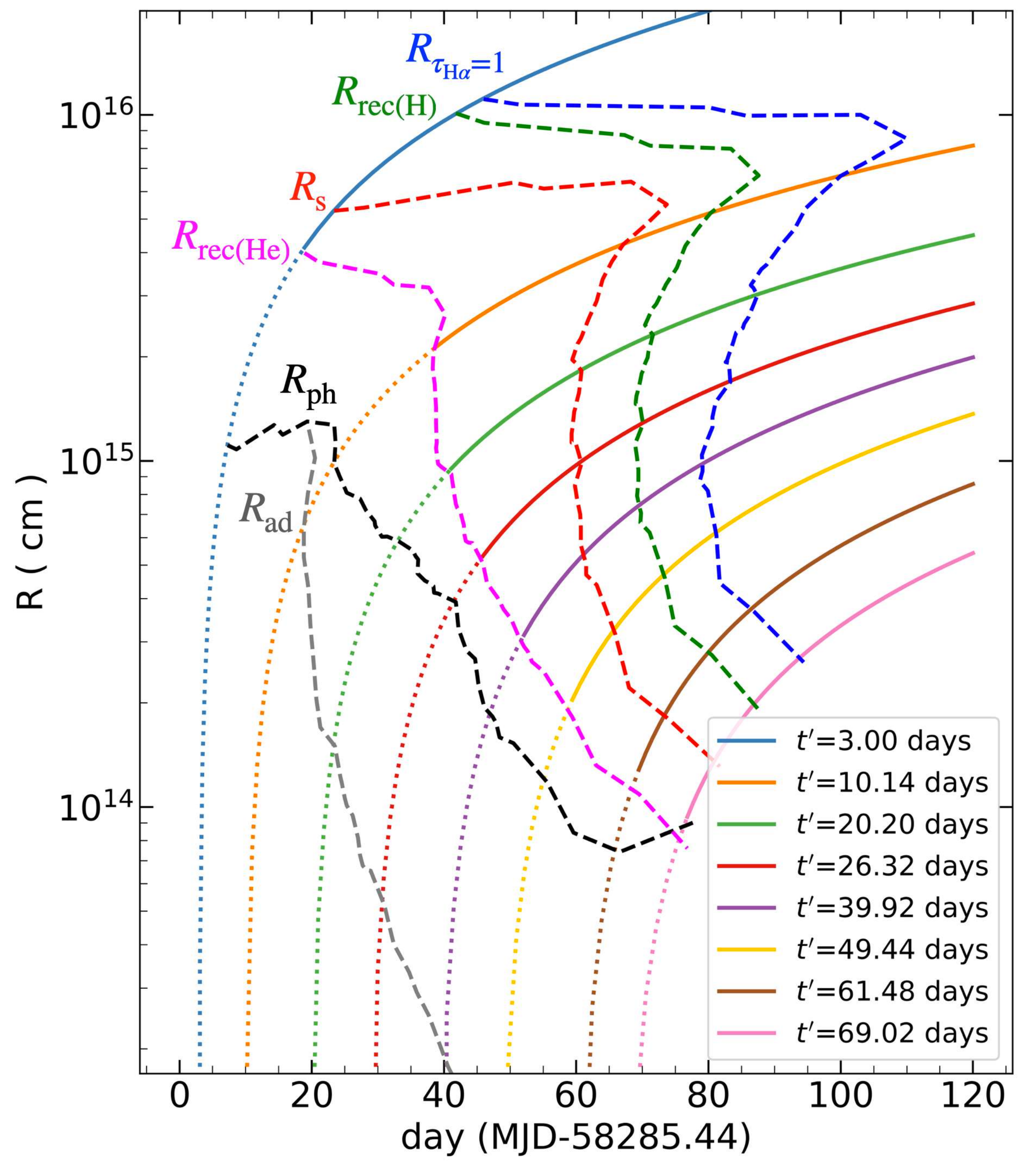}
\caption{The histories of selected Lagrangian fluid elements, and the characteristic physical scales in the model for AT2018cow. The zero point in the y-axis is set at $R_{\rm eq}$ ($1.7\times 10^{13}{\rm ~cm}$). At $t\lesssim 20$ days, the color radius $(R_{\rm c})$ is below $R_{\rm ad}$, and thus $R_{\rm ph} = R_{\rm ad}$. At $t\gtrsim 20$ days, the photosphere is formed at $R_{\rm c}$ (i.e., $R_{\rm ph} = R_{\rm c}$). The dotted and solid lines show a track of each Lagrangian fluid element, $R = v(t-t^{\prime}) + R_{\rm eq}$, where $t^{\prime}$ is the time when each Lagrangian fluid element is launched and it is shown in the labels. The dashed lines show some characteristic radii.}
\label{fig:fig4}
\end{figure}

The model has a monotonically decreasing velocity evolution. Therefore, the outflows launched at later epochs never catch up with those ejected at earlier epochs. This means that the steady-state solution is a good approximation, as long as the effect of the infinite time delay is taken into account. Here, the time delay means a time interval for each Lagrangian fluid element to experience between the launch ($R_{\rm eq}$) and the arrival at the photosphere ($R_{\rm ph}$). This can be partly accounted for, by examining a history of each Lagrangian fluid element. The history of each Lagrangian fluid element is shown in Figure \ref{fig:fig4}, which also shows characteristic physical scales (e.g., $R_{\rm ad}$) encountered by each Lagrangian fluid element. Indeed, Figure \ref{fig:fig4} shows that the time delay is sufficiently small. Therefore our procedure to estimate $\dot{M}$ and $v$ from the observational data at the photosphere ($L$ and $T_{\rm ph}$) without including this time delay would not introduce a large error.

\begin{figure*}
\epsscale{1.17}
\plotone{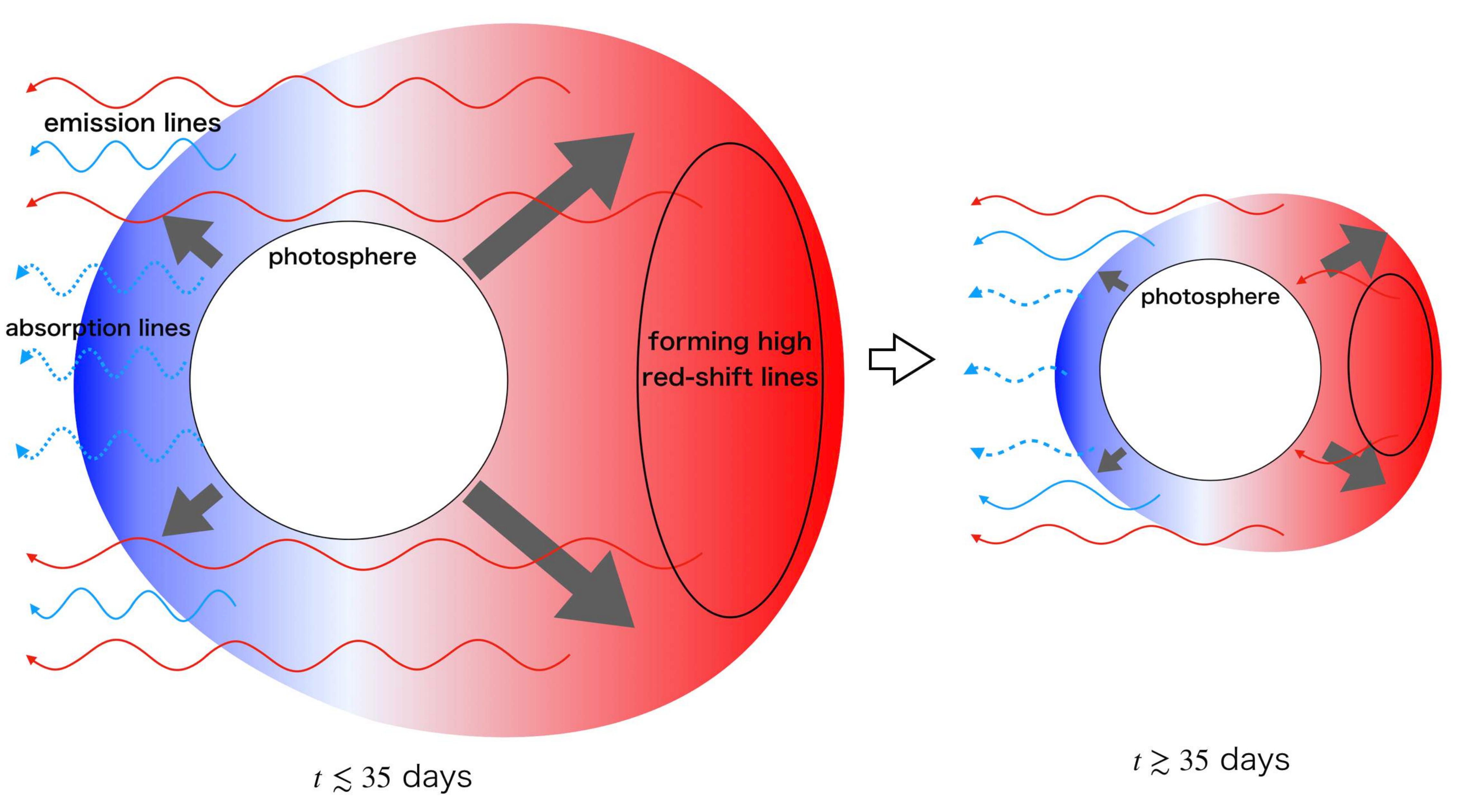}
\caption{A schematic picture (not scaled) for the line formation in AT2018cow, where an observer is placed on the left side of the figure. It is assumed that the wind in the right side is stronger, to explain the redshift observed in the early phase. The change in the relative size of the photosphere to the recombination radius results in the change in the line profile (see the main text). (Left: ) At $t \lesssim 35$ days, the photospheric radius is much smaller than the recombination radius of helium. (Right: ) At $t \gtrsim 35$ days, both of the photosphere and the recombination front move inward. The shrink of the recombination radius is more substantial, leading to the large photospheric radius relative to the recombination radius.}
\label{fig:fig5}
\end{figure*}

Figure \ref{fig:fig4} allows to extract general features in spectral line formation expected for this model. Using the recombination temperature of helium, $T_{\rm rec(He)} \approx 12000{\rm ~ K}$, we can derive the recombination radius of helium, $R_{\rm rec(He)}$, for each Lagrangian fluid element, below which helium is singly ionized and creates no \ion{He}{1} lines by resonance scattering. The initial outflow $(v\approx 0.1c)$ injected from $R_{\rm eq}$ at $\sim 3$ days approaches $R_{\rm rec(He)}$ on $\sim 20$ days (see Figure \ref{fig:fig4}). Given that it takes $\sim 5$ days for the initial wind Lagrangian fluid element to reach to $R_{\rm ph}$ and start emitting photons, the wind-driven model predicts that the \ion{He}{1} lines start to emerge $\sim 10-15$ days after the discovery. This result is consistent with the observation \citep{2019MNRAS.484.1031P} which shows the emergence of the \ion{He}{1} lines at $\sim 15$ days.

The recombination temperature of hydrogen, $T_{\rm rec(H)}$, is taken as $6000{\rm~ K}$. Similarly to the case for the helium recombination, hydrogen is fully ionized below $R_{\rm rec(H)}$. The hydrogen line forming region, $R_{\rm \tau_{\rm H\alpha} = 1}$ (see Appendix \ref{sec:appendixA}), closely follows $R_{\rm rec(H)}$ up to $\sim 60$ days. The epoch we estimate for the hydrogen lines to emerge is therefore $\sim 40$ days, which is later than what is seen in the observation by a factor of two. However, we note that the temperature decrease will be accelerated, once additional cooling effect is considered. Especially, the helium recombination would cool the outflow efficiently, which might decrease $R_{\rm rec(H)}$ and $R_{\tau_{\rm H\alpha =1}}$, leading to the formation of the H lines immediately after the He line formation.

The hydrogen line forming radius, $R_{\tau_{\rm H\alpha =1}}$, is larger than $R_{\rm ph}$ by more than an order of magnitude. Even if we assume that the hydrogen lines would be formed at $R_{\rm rec(He)}$ (see above), it is so by a factor of $\gtrsim 3$ in the first $\sim 40$ days. When the line forming region is far above the photosphere, the spectra must be characterized by emission lines (see Figure \ref{fig:fig1}). This result is consistent with the observed spectra of AT2018cow, which show emission lines, not absorption.

At $t\gtrsim 60$ days, Figure \ref{fig:fig4} shows that $R_{\rm s}$, $R_{\rm rec(H)}$, and $R_{\rm rec(He)}$ have multiple values for a given epoch. This stems from the decreasing mass-loss rate, leading to the smaller characteristic radii for the Lagrangian fluid elements launched at later epochs. This behavior leads to a situation where a similar temperature is realized in a wide spatial range. However, having multiple values in the characteristic radii, which would lead to a complicated neutral-ionized-neutral structure, is likely an artifact. In reality, the inversed temperature structure will be smeared out by radiation diffusion.

The observed hydrogen and helium lines show redshifts of $\sim 3000{\rm ~km~s^{-1}}$ at the time of the first detection of the lines. They evolve blueward as time goes by, and change the profile at $\sim 30-40$ days, after which they show sharp peaks around the rest wavelengths, with the bluer flux suppressed \citep{2019MNRAS.484.1031P}. This behavior is explained naturally within a context of the wind-driven model (Figure \ref{fig:fig5}). The possible explanation of the (initial) redshift here is phenomenological, but it can be explained if we consider aspherical winds and we observed the event from the `weaker' side. The peculiar time evolution is, on the other hand, predicted by our wind model irrespective of the wind geometry (note that the spectral information is not used in constructing the model). Figure \ref{fig:fig4} shows that the recombination radius of helium is substantially larger than the photospheric radius until $\sim 35$ days after the discovery. In this phase, the emission line is expected, and the line profile follows the geometrical distribution of the wind. Around $\sim 35$ days, the recombination radius of helium suddenly decreases and becomes close to the photospheric radius. Afterward, the red-shifted emission from the rear region are efficiently blocked by the photosphere, and the profile we observe should evolve blueward. In addition, the line profile is affected by the absorption for the approaching side, and the blue-shifted emission component will be suppressed by this effect. Therefore, we expect to observe a sharp profile at the rest-frame wavelengths, with the blue-shifted side substantially suppressed.

\begin{figure}
\epsscale{1.17}
\plotone{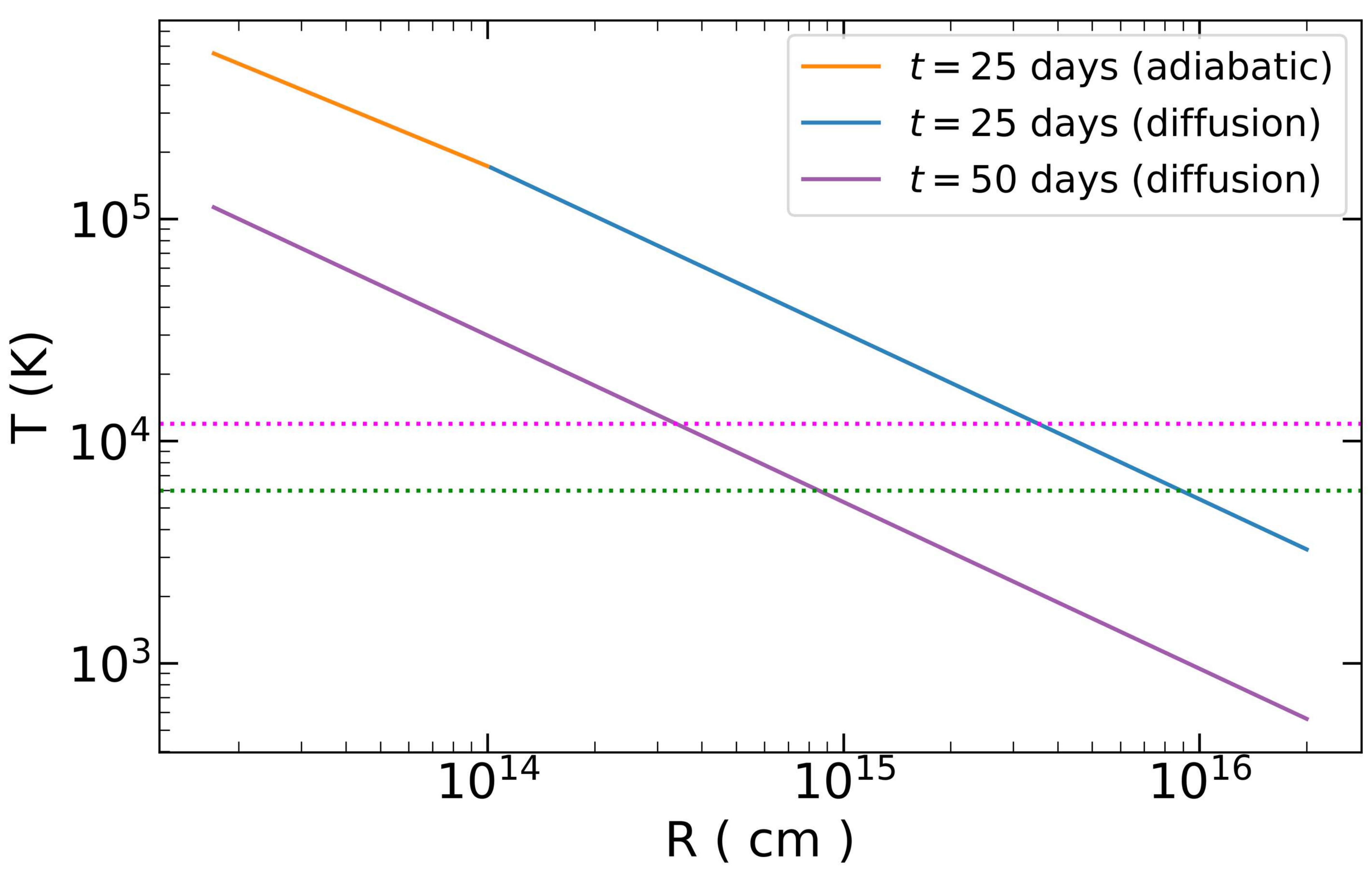}
\caption{The temperature structures at $25$ and $50$ days in the steady-state solution (see the main text). The magenta dotted line shows the He recombination temperature $(T_{\rm rec(He)} \approx 12000{\rm ~K})$, and the green dotted line shows the H recombination temperature $(T_{\rm rec(H)} \approx 6000{\rm ~K})$.
}
\label{fig:6}
\end{figure}

Figure \ref{fig:6} shows the temperature structures at $25$ and $50$ days in the steady-state solution (i.e., without convolution of contributions from different Lagrangian fluid elements). At 25 days, a break in the temperature structure is seen at $R_{\rm ad} \sim 10^{14}{\rm ~cm}$, with the power law index of $-2/3$ (adiabatic) and $-3/4$ (diffusion), below and above $R_{\rm ad}$, respectively. At 50 days, the structure follows a single power law with the index of $-3/4$, since the region immediately above $R_{\rm eq}$ is already in the diffusion dominated region at this phase. As shown in Figure \ref{fig:6}, the temperature in the wind model for AT2018cow is overall decreasing with time.

\subsection{iPTF14hls} \label{subsec:3.2}

\begin{figure}
\epsscale{1.17}
\plotone{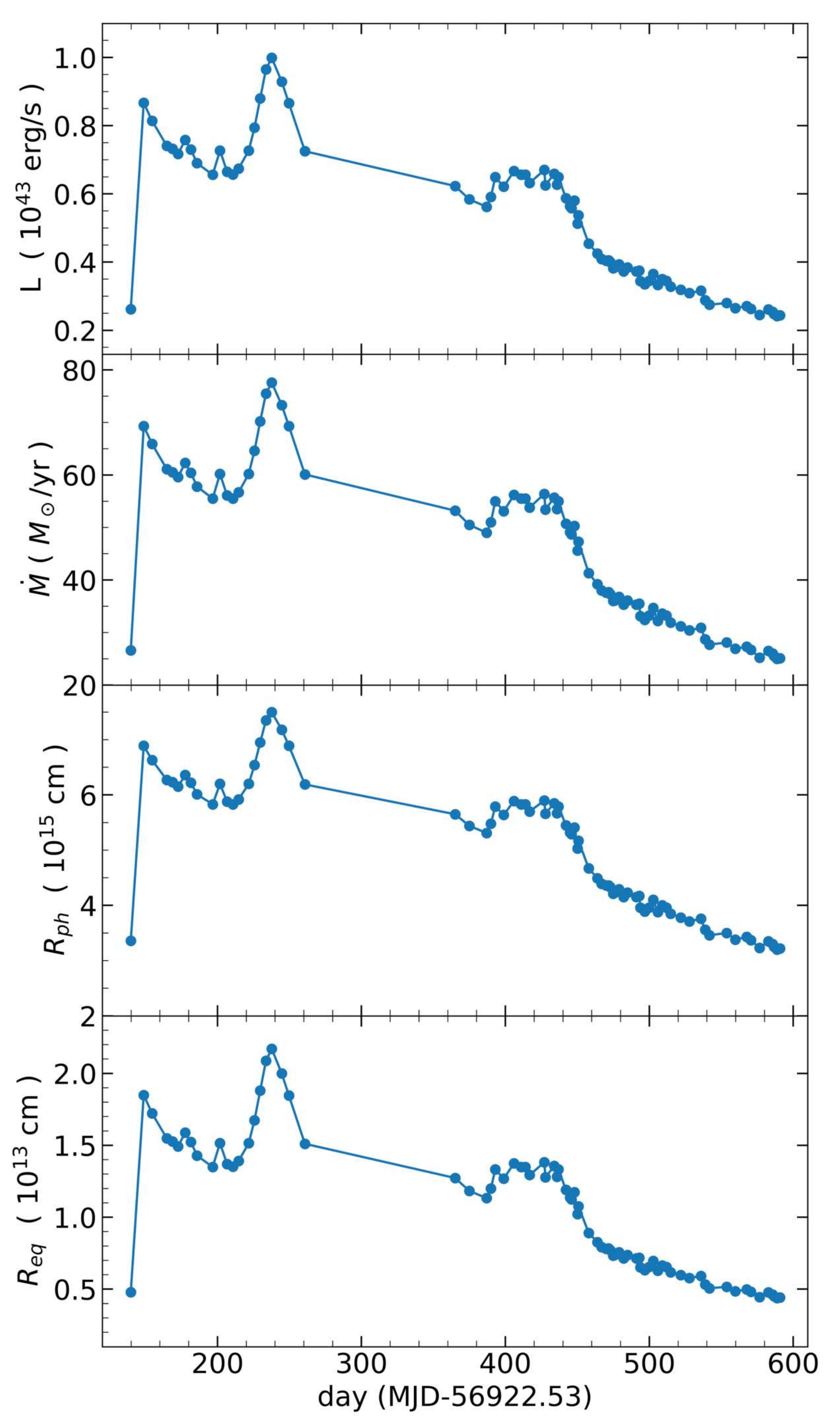}
\caption{The top panel shows the bolometric light curve of iPTF14hls from \cite{2017Natur.551..210A}. The second panel shows the evolution of the derived mass-loss rate. The third panel shows the estimated photospheric radius evolution. The bottom panel shows the evolution of the estimated inner radius.}
\label{fig:fig7}
\end{figure}

For iPTF14hls, $T_{\rm ph} \sim 7250{\rm ~K}$ has been derived, which does not evolve much over time \citep{2020MNRAS.491.1384M}. This is close to the recombination temperature of hydrogen ($T_{\rm rec(H)}\approx 6000{\rm ~K}$). Therefore, the effect of the recombination radius must be taken into account. 
In addition, the velocity of \ion{Fe}{2} lines stayed nearly constant, $v \approx 4000{\rm ~km~s^{-1}}$, over time \citep{2017Natur.551..210A}. This velocity should represent the outward velocity around the photosphere. The number of the observational constraints is enough to derive a unique solution (equations \ref{eq:close1}, \ref{eq:close2}, and \ref{eq:close3}). For the conditions appropriate for iPTF14hls, it turns out that $R_{\rm c}$ is always larger than $R_{\rm ad}$, therefore $R_{\rm ph} = R_{\rm c}$ and $T_{\rm ph} = T(R_{\rm c})$. Accordingly, $\dot{M}$ is described as follows;
\begin{equation}
  \begin{split}
    \dot{M}(t) = 77.&6 M_{\odot}{\rm ~yr^{-1}}\\
    &\times\left(\frac{v(t)}{4.00\times10^8{\rm ~cm~s^{-1}}}\right)\\
    &\times\left(\frac{L(t)}{1.00\times10^{43}{\rm ~erg~s^{-1}}}\right)^{\frac{4}{5}}.
  \end{split}
\end{equation}
Figure \ref{fig:fig7} shows the evolution of $\dot{M}$, $R_{\rm ph}$, and $R_{\rm eq}$ as we have derived.

\begin{figure}
\epsscale{1.17}
\plotone{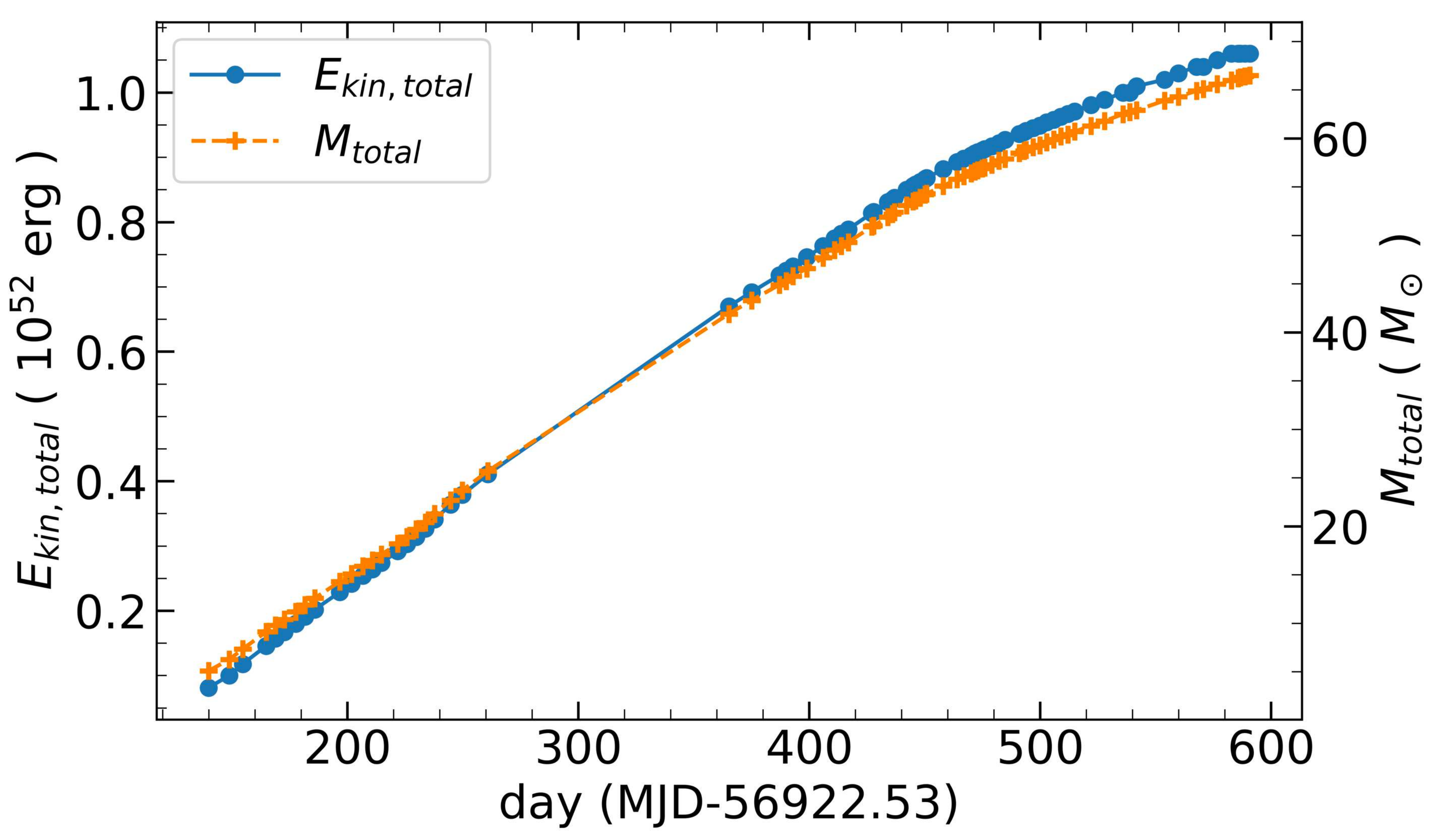}
\caption{The same as Figure \ref{fig:fig3} but for iPTF14hls.}
\label{fig:fig8}
\end{figure}

At the maximum luminosity, the mass-loss rate in the model is over $75M_{\odot}{\rm ~yr^{-1}}$. The large mass-loss rate here is qualitatively consistent with that suggested by \citet{2020MNRAS.491.1384M} based on a phenomenological approach where they assumed the density at the photosphere, while we derive it by using other constraints. Indeed, the quantitatively derived mass-loss rate in this work is larger by a factor of $\sim 3$. To explain the observational properties of iPTF14hls, the outflows (winds) should keep its strength for almost 2 years. The total ejected mass is $\sim 66M_{\odot}$ and the total kinetic energy is $\sim 1.0\times 10^{52}{\rm ~erg}$ in the wind-driven model (see Figure \ref{fig:fig8}).

\begin{figure}
\epsscale{1.17}
\plotone{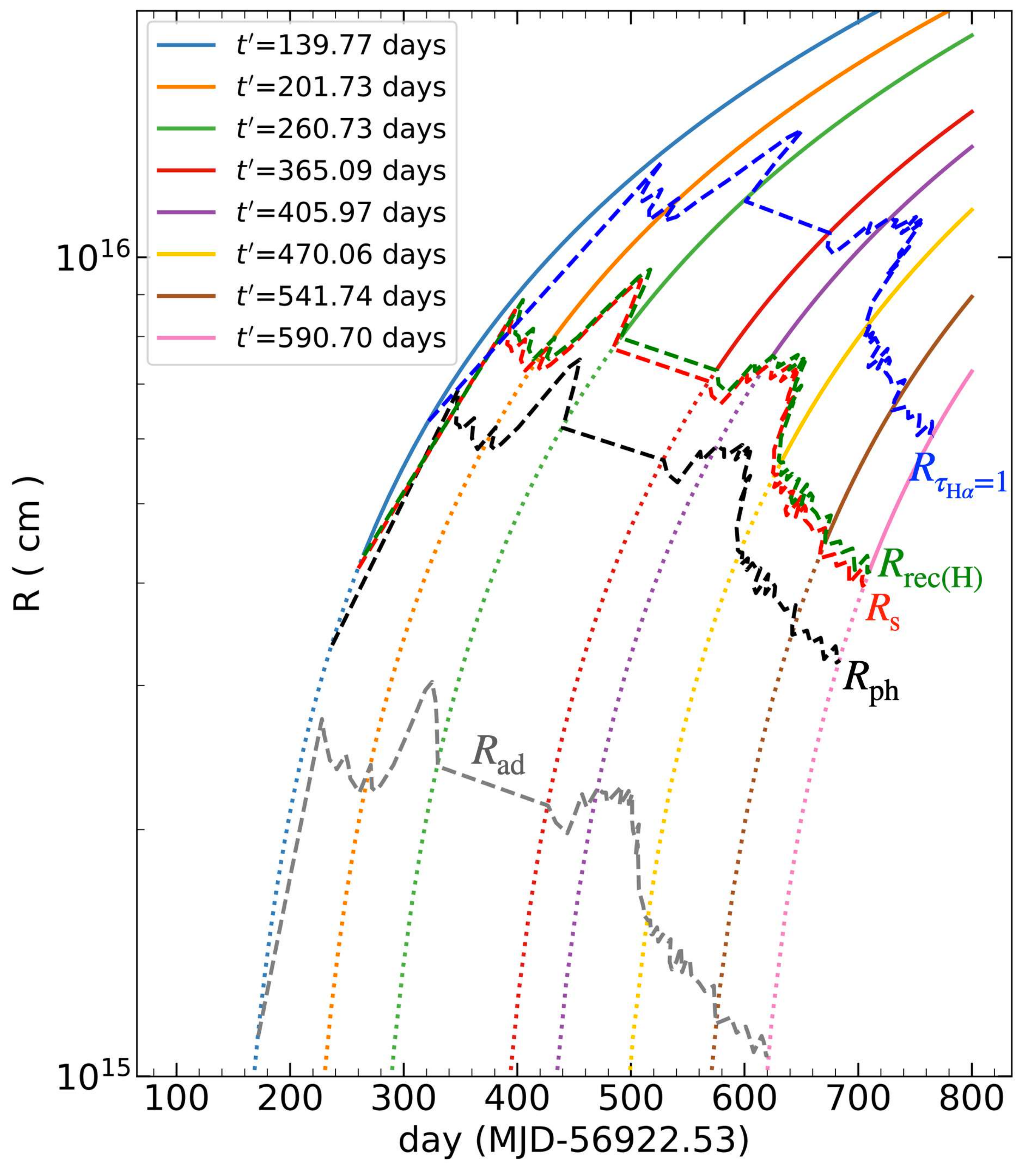}
\caption{The same as Figure \ref{fig:fig4} but for iPTF14hls. For presentation, the zero point in the y-axis is not set at $R_{\rm eq}$.}
\label{fig:fig9}
\end{figure}

Figure \ref{fig:fig9} shows the evolution of some characteristic radii, overplotted with the histories of selected Lagrangian fluid elements. Below $R_{\rm rec(H)}$, hydrogen is fully ionized. The hydrogen recombination occurs when the Lagrangian fluid element reaches to $R_{\rm rec(H)}$. For iPTF14hls, typical time delay for each Lagrangian fluid element to move from $R_{\rm eq}$ to $R_{\rm ph}$ (and $R_{\rm \tau_{\rm H\alpha} = 1}$) is $\sim 100$ days. Given the overall slow evolution of iPTF14hls until $\sim 450$ days, the steady-state approximation is justified. 
Note that our model applies only after $\sim 250$ days in Figure \ref{fig:fig9} in discussing the spectral properties. Namely, the initial $\sim 140$ days delay in the emergence of the hydrogen lines inferred in Figure \ref{fig:fig9} is an artifact, since our model is constructed only through the multi-color observational data after $\sim 140$ days since the discovery. If we assume the same time delay in the emergence of the hydrogen lines for the earlier epochs, it would predict that iPTF14hls must have had the hydrogen lines emerged at $\sim 110$ days since the initiation of the explosive event. It is consistent with the presence of the P-Cygni hydrogen lines in the first spectrum reported, taken at 104 days after the discovery \citep{2017Natur.551..210A}.

Considering the Sobolev approximation (see Appendix \ref{sec:appendixA}) as we have done for AT2018cow, we estimate the line-forming radius for ${\rm H\alpha}$ $(R_{\tau_{\rm H\alpha =1}})$. We find that $R_{\tau_{\rm H\alpha =1}}$ is larger than $R_{\rm ph}$ by only a factor of at most two (note that, for AT2018cow, $R_{\tau_{\rm H\alpha =1}}$ is at least by a factor of $\sim 3$ or even more than an order of magnitude larger than $R_{\rm ph}$). In this case, we expect to observe the P-Cygni profiles (see Figures \ref{fig:fig1} and \ref{sec:5}). iPTF14hls did show the P-Cygni profiles up to $\sim 450$ days, fully consistent with our result. We emphasize that this spectral information is not used in constructing the wind-driven model, and this spectral behavior is a model prediction (note that \citet{2020MNRAS.491.1384M} qualitatively estimated the density at the photosphere, using the spectral information). 

\begin{figure}
\epsscale{1.17}
\plotone{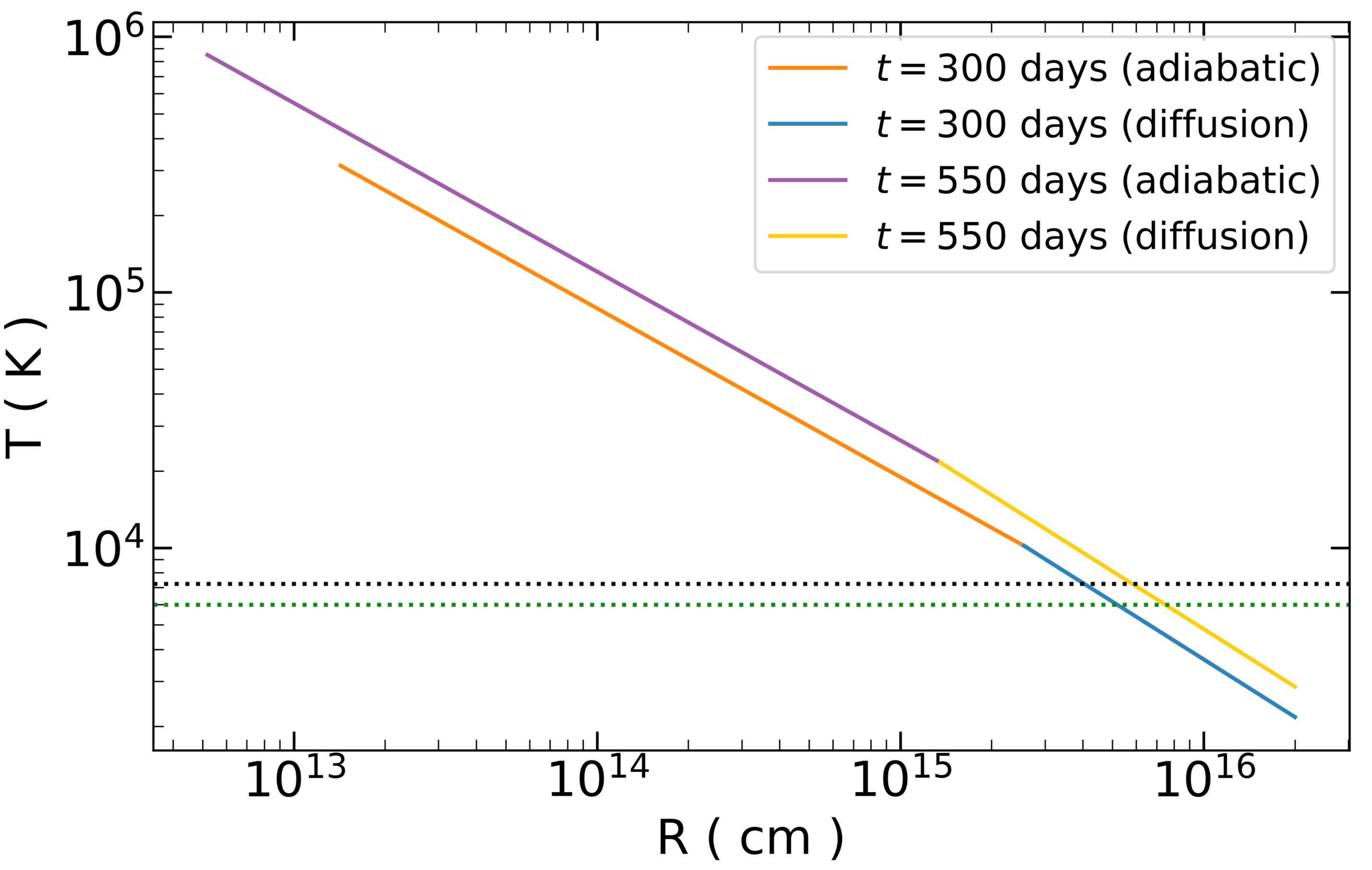}
\caption{The same as Figure \ref{fig:6} but for iPTF14hls. The black dotted line shows the observed photospheric temperature $(T_{\rm ph}\sim 7250{\rm ~K})$.}

\label{fig:10}
\end{figure}

Figure \ref{fig:10} shows the temperature structures at $300$ and $550$ days in the steady-state solution. The inner radius $(R_{\rm eq})$ moves inward, and $T_{\rm eq}$ becomes higher. However, the change in the temperature structure is not large, reflecting the slow evolution of iPTF14hls.

\section{DISCUSSION} \label{sec:4}

In this paper, we have shown that peculiar properties of AT2018cow and iPTF14hls, which have not been explained by the existing models like a supernova explosion, can be naturally explained by the wind-driven model. Furthermore, although AT2018cow and iPTF14hls have very different observational properties, we have shown that they can be explained within the same context of the wind-driven model.
Interestingly, both events have almost the same inner radii, $\sim 10^{13}{\rm ~cm}$. On the other hand, the main differences in the derived properties are their kinetic energies, total ejected masses, and time scales.

The physical scale where the equipartition takes place, $R_{\rm eq} \sim 10^{13}{\rm ~cm}$, is a typical radius of a red super giant (RSG). This result implies that the progenitor (system) may involve an RSG. The energy budget, $\sim 10^{51-52}{\rm ~erg}$, indicates that it may be powered by the release of the gravitational energy at $\sim 10^{10-11}{\rm ~cm}$, if this is powered by a stellar object (i.e., $10-100M_{\odot}$). Interestingly, this is the size of a core of an RSG. Except for SNe, phenomena which could release such a large amount of kinetic energy are limited.

As one possibility, we consider a binary system including an RSG, specifically the mass ejection driven by a common envelope (CE) evolution as the energy source. Given the mass ejection of $\sim 66M_{\odot}$ in iPTF14hls, we may consider a CE where the primary's He core mass is $\sim 50 M_{\odot}$ (i.e., $\sim 120M_{\odot}$ as a whole) and a companion star is $\sim 100M_{\odot}$.

The typical dynamical time scale (free-fall time) of the primary's envelope is given by
\begin{align}
\label{free-fall}
    t_{\rm dyn} &= \sqrt{\frac{3\pi}{32G\rho}} \nonumber\\
    &\sim 10{\rm ~ days}\left(\frac{M}{100M_{\odot}}\right)^{-\frac{1}{2}}\left(\frac{R}{2\times 10^{13}{\rm ~cm}}\right)^{\frac{3}{2}},
\end{align}
where $G$ is the Newtonian constant of gravitation. The dynamical time scale shown here would set a minimal response time in which the mass ejection reacts to the change in the energy input from the central system (a merged core or a close binary within an RSG envelope). This would then give the typical time scale of the variability in its luminosity. Indeed, the typical time scale of the variability seen in the light curve of iPTF14hls ($\sim 10$ days) is roughly on the same order.

If the orbital separation between the core and the companion shrinks to $\sim 10^{11}{\rm ~cm}$ \citep[i.e., the core size, see also ][]{2019ApJ...878...49W}, the orbital energy release $(E_{\rm grav})$ is estimated as follows;
\begin{align}
  \begin{split}
    E_{\rm grav} \approx 1.3&\times 10^{52}{\rm ~ erg} \\ 
    &\times \left(\frac{M_{1,\rm core}}{50M_\odot}\right)\left(\frac{M_2}{100M_\odot}\right)\\
    &\times\left(\frac{R_{1,\rm core}}{1\times10^{11}{\rm~cm}}\right)^{-1},
  \end{split}
\end{align}
where $M_{1,\rm core}$ is the primary's core mass, $R_{1,\rm core}$ is the core radius, and $M_{2}$ is the companion mass. This roughly explains the estimated total kinetic energy for iPTF14hls. 

Candidates for the primary and companion stars are population I\hspace{-.1em}I\hspace{-.1em}I or low-metallicity stars. They have a few $R_{\odot}$ and $\sim 100M_{\odot}$ at the zero age main sequence \citep{2012A&A...542A.113Y}. Assuming a binary with initial masses of $120M_{\odot}$ and $100M_{\odot}$, the primary should evolve to an RSG first, and its hydrogen envelope $(\sim 70 M_{\odot})$ expands and fills the Roche lobe. Then, the Roche lobe overflow (RLOF) is likely unstable, leading to the CE mass ejection and the merge of the primary's He core and the companion. Note that the release of the gravitational energy at $\sim 10^{12}{\rm ~cm}$ is enough to unbind the hydrogen envelope whose binding energy is $\sim 10^{51}{\rm ~erg}$. However, details of a CE evolution are not yet clarified; if the time scale of the orbital decay is shorter than the mass ejection, the separation would further decrease, and the core merger would take place.

A similar scenario may also be realized for a binary of a massive He star and an RSG, in which the physical scale of $\sim 10^{11}{\rm ~cm}$ will be set by the sizes of the He star and the He core. This may also happen for a massive binary (initially $\sim 100M_{\odot}$ for each component), if the first RLOF mass transfer is stabilized by the inversed mass ratio shortly after the initiation of the RLOF. The second RLOF should then be unstable, similar to the above scenario. In this scenario, a low metallicity may not be required to have the physical scale of $\sim 10^{11}{\rm ~cm}$ (since this is set by the sizes of the He star/core). However, to keep a large amount of the hydrogen envelope through the evolution, the low-metallicity condition is required. Yet another scenario would be a merger of two RSGs as discussed by \citet{2019ApJ...884...58S}.

In any case, the above mentioned evolutionary scenarios are largely speculative. Our main focus in the present work is to show the applicability of the wind-driven model for iPTF14hls (and AT2018cow), and how to realize the inferred initial condition is beyond a scope of the paper. In addition, details of the CE interaction and the wind launch require further investigation in the future; for example, nuclear reactions of the merged core would be enhanced, and then a less massive binary system than considered here may satisfy the energy budget requirement. In summary, as one possibility we suggest the dynamical CE evolution induced by a massive binary system ($\sim 100M_{\odot}$ each) as a possible scenario for iPTF14hls.

This CE scenario is, however, not suitable to AT2018cow. The dynamical time scale of a putative RSG companion is too long for AT2018cow (see the equation \ref{free-fall}); given the smaller mass ejection, we might consider a less massive companion RSG which leads to even larger time scale. Furthermore, the evolution of the mass-loss rate ($\propto t^{-5/3}$) suggests that it is probably driven by a fallback accretion onto a BH. Also, the fast ejecta ($\sim 0.1c$) indicates an event related to a compact object.

We suggest two scenarios that could satisfy these constraints; a BH-forming failed SN or a TDE of an RSG. For the BH-forming failed SN of a massive RSG, only the outermost layer, thus $\lesssim 1M_{\odot}$, is ejected \citep{2015MNRAS.451.2656K}. The energy scale of the fallback accretion is given by $E = \epsilon M c^2$, where $\epsilon \sim 10^{-3}$ \citep{2013ApJ...772...30D}. If we consider $1M_{\odot}$ as the accreted mass, then it is $\sim 1-2 \times 10^{51}{\rm ~erg}$. Another possibility is a TDE of a low-mass RSG. The energy budget will be similar to the case of the failed-SN scenario.

The fallback accretion scenario for AT2018cow is similar to the suggestion by \citet{2019ApJ...872...18M}. Their suggestion is based on qualitative analyses of the multi-wavelength data including X-ray observations, while we here focus on the quantitative interpretation of the optical/UV data. The analyses are thus complementary. Indeed, AT2018cow shows unique features not only in the optical/UV range but also across the wavelengths, including radio and X-ray emissions. As future work, we plan to extend the present model to provide quantitative prediction in the other wavelengths.

\section{CONCLUSIONS} \label{sec:5}

The peculiar transients AT2018cow and iPTF14hls showed unique observational properties, a combination of which defies straightforward explanations by existing models (e.g., an SN-like explosion). AT2018cow showed a rapidly decreasing luminosity and a recessing photosphere. iPTF14hls showed a long-lasting luminosity for almost 2 years, constant line velocities, and too slow spectral evolution. In the present work, we have proposed a model, the wind-driven model, to explain these two peculiar transients with totally different observational features. We have shown that the model can explain the light curves and spectral evolution for both of AT2018cow and iPTF14hls.

Under the wind-driven model, we have estimated the evolution of the mass-loss rate, $\dot{M}$. Both transients are explained by (initially) strong outflows exceeding a few $M_{\odot}{\rm ~yr^{-1}}$ ($\sim 20M_{\odot} {\rm ~yr^{-1}}$ for AT2018cow and $\sim 75M_{\odot} {\rm ~yr^{-1}}$ for iPTF14hls). In addition to this similarity in the mass-loss rates, they share the innermost (equipartition) radius of $\sim 10^{13}{\rm ~cm}$. On the other hand, their kinetic energies, total ejected mass, and time scales are different.

The model does not use the information on the spectral line features in its construction. Therefore, we can provide `prediction' for the spectral features. We have shown that the model can explain the characteristic spectral feature; emission in AT2018cow while absorption (or P-Cygni) in iPTF14hls. We can also explain the evolution and related time scales seen in AT2018cow; emergence of \ion{He}{1} lines at $\sim 15$ days, the blueward shift toward the rest wavelength in the red component, as well as suppression in the blue wing, in time scale of $\sim 30$ days.

The radius of $\sim 10^{13}{\rm ~cm}$ suggests that both events likely involve an RSG. The kinetic energy of $\sim 10^{51-52}{\rm erg}$ then matches to the gravitational energy release if the system would shrink to $\sim 10^{11}{\rm cm}$. This is the typical size of a He core of an RSG, and we speculate this may be related to a common-envelope event involving an RSG as a primary for iPTF14hls, in a low-metallicity massive binary system. AT2018cow has a much shorter time scale than iPTF14hls, and we speculate that the companion star here is a BH. This can then be a TDE involving a low-mass RSG, or a BH-forming failed SN from a massive RSG.

In the present work, we have restricted ourselves for the steady-state solution (see \citet{2020ApJ...894....2P} for discussion of the effect of a non-steady-state wind). While we have shown that it is a good approximation and also have taken into account the effect of the time delay in the spectral formation analysis, detailed and accurate investigation will require radiation-hydrodynamic simulations. Also, spectral synthesis simulations are required to address further details of the spectral evolution. We plan to tackle these issues in our future work.

\acknowledgments

K.M. acknowledges support provided by Japan Society for the Promotion of Science (JSPS) throng KAKENHI grant (17H02864, 18H04585, 18H05223, 20H00174, and 20H04737).

\appendix

\section{Sobolev approximation} \label{sec:appendixA}

We use the Sobolev approximation \citep{1960mes..book.....S, 1970MNRAS.149..111C} to compute the line optical depth, neglecting the stimulated emission. In general, it is given as follows;
 \begin{align}
  \tau_{\nu_{0}}=\frac{\pi e^{2}}{m_{\rm e} c} f_{l} \lambda_{\nu_{0}} n_{l}\frac{1}{\left\{\frac{d v}{d r} \cos ^{2} \theta+\frac{v(r)}{r}\left(1-\cos ^{2} \theta\right)\right\}},
 \end{align}
where $m_{\rm e}$ is the mass of electron, $f_{l}$ is the line oscillation strength, $n_{u}$ and $n_{l}$ is the number density in the upper level and the lower level, $\lambda_{\nu_{0}}$ is the line rest wavelength, and $\theta$ is the between the flow direction and the line of sight.

For $\rm H\alpha$, $n_{l} = n_{2}$, where $n_{2}$ is the number density of hydrogen in the second level. For the steady state wind, $dv/dr = 0$. We could estimate the line optical depth by setting $\theta = 90^{\circ}$, $f_{2} \approx 0.64$ and $\lambda_{\rm H\alpha} = 656.3 {\rm ~nm}$.
To estimate $n_{2}$, we use the density and temperature computed for the wind-driven model. Assuming $n_{1} \approx n_{\rm H}$ and the Boltzmann distribution, where $n_{\rm H}$ is the number density of hydrogen, and $n_{1}$ is that of hydrogen in the ground level, $n_{2}$ is given by
\begin{align}
 n_{2} \approx \frac{Y_{\rm H}}{\mu m_{\rm p}}\rho \frac{g_{2}}{g_{1}}\exp\left({-\frac{\Delta E_{1,2}}{kT}}\right),
\end{align}
where $Y_{\rm H} \approx 0.9$ is the number fraction of hydrogen for the solar composition, $\mu \approx 1.34$ is the mean atomic mass, $m_{\rm p}$ is the proton mass, $g_{1}=2$ and $g_{2}=8$ are the statistical weights, $E_{1,2} = 10.2{\rm ~eV}$ is the energy difference, and $k = 8.62\times10^{-5}{\rm ~eV~K^{-1}}$ is the Boltzmann constant. 
Therefore, the line optical depth of $\rm H\alpha$ is derived as follows;
\begin{equation}
  \begin{split}
    \tau_{\rm H\alpha} &\approx 1.79\times 10^{18}~ \rho \exp\left({-\frac{\Delta E_{1,2}}{kT}}\right) \frac{r}{v},
  \end{split}
\end{equation}
where $\rho$, $r$, and $v$ are expressed in the cgs unit. The line forming radius is evaluated by $\tau_{\rm H\alpha}\approx 1$.

\bibliography{manuscript}{}
\bibliographystyle{aasjournal}

\end{document}